# Observation of abnormal resistance-temperature behavior along with diamagnetic transition in $Pb_{10-x}Cu_x(PO_4)_6O$-based composite

Hao Wu[1,2]†, Li Yang[1,2]†, Jie Yu[1,2], Gaojie Zhang[1,2], Bichen Xiao[1,2], Haixin Chang[1,2]*

[1]State Key Laboratory of Material Processing and Die & Mold Technology, School of Materials Science and Engineering, Huazhong University of Science and Technology (HUST), Wuhan 430074, China.

[2]Wuhan National High Magnetic Field Center and Institute for Quantum Science and Engineering, Huazhong University of Science and Technology (HUST), Wuhan 430074, China.

†These authors contributed equally to this work.

*Corresponding author. E-mail: hxchang@hust.edu.cn

**Abstract**

Recently, Sukbae Lee et al. reported that material $Pb_{10-x}Cu_x(PO_4)_6O$ (LK-99) has a series of characteristics of room temperature superconductors, including diamagnetic transition, resistance jump, nearly zero-resistance, magnetic field-dependent IV characteristics and so on (10.6111/JKCGCT.2023.33.2.061, arXiv:2307.12008, arXiv:2307.12037). However, whether LK-99 is really a room temperature superconductor is still controversial. On the one hand, some people think that the relatively weak diamagnetism of LK-99 reported by Sukbae Lee et al. is not the Meissner effect. On the other hand, there are doubts about the authenticity of its zero-resistance test results. Global replication studies have shown that LK-99 does have a large diamagnetic (arXiv:2308.01516), and also found a zero-resistance behavior at a low temperature of 110 K (arXiv:2308.01192). However, up to now, there is still no direct reproducible evidence to support Sukbae Lee et al. 's conclusion that LK-99 is a room temperature superconductor. Here, a distinct resistance jump was observed at about 387 K under ambient pressure in our experiment for unclear reason including possible impurity's contribution. The overall resistance of the test LK-99 sample still shows semiconductivity, and the resistance cannot really drop to zero. Our findings indicate that to identify the true potential of LK-99, high quality crystals without impurity are very important.

**Introduction**

The research of superconducting materials has brought great influence on the fields of science and engineering, such as magnetic resonance imaging, high-energy particle accelerator and energy transmission. In 1986, J. G. Bednorz et al. discovered the first high-temperature superconductor in the La-Ba-Cu-O system, which caused a stir in the academic community.[1] However, the highest $T_c$ of current atmospheric-pressure superconducting materials is only 133 K,[2] which is far lower than the room temperature of 300 K, and the practical application still faces high low-temperature costs. In addition, A. P. Drozdov et al. found that sulfur hydride will undergo superconducting phase transition at a temperature of about 203 K in an extremely high-pressure environment (at least 150 GPa, that is, about 1.5 million standard atmospheres).[3] However, compared with the low-temperature cost, the high-pressure cost limits its practical application. More recently, Sukbae Lee et al. reported the LK-99 material with a $T_c$ higher than the boiling point of water at atmospheric pressure, with obvious diamagnetic transition, nearly zero-resistance, and obvious magnetic levitation effect.[4-6] However, to date, the room temperature superconductivity of LK-99 has not been fully confirmed experimentally.

**Experiment**

In this manuscript, LK-99 samples were grown by solid-state method with lead oxide yellow (PbO, 99.999%, Aladdin), lead sulfate (PbSO$_4$, 99.99%, Aladdin), phosphorus red (P, 99.999%, Aladdin) and copper ultrafine powder (Cu, 99.9%, Macklin) as source materials. Firstly, to synthesize Cu$_3$P, Cu powder and P with a molar ratio of 3:1 were thoroughly mixed and ground, and sealed in a quartz ampoule under a vacuum of 10$^{-2}$ Pa. Then the mixture was heated to 550 °C within 2 h and kept for 48 h to obtain the precursor Cu$_3$P. Subsequently, to synthesize Lanarkite (Pb$_2$(SO$_4$)O), PbO and PbSO$_4$ with a molar ratio of 1:1 were thoroughly mixed and ground, and then sealed in a quartz ampoule under a vacuum of $10^{-2}$ Pa. Then the mixture was heated to 725 °C within 2 h and kept for 24 h to obtain the precursor $Pb_2(SO_4)O$. Finally, to synthesize $Pb_{10-x}Cu_x(PO_4)_6O$, $Cu_3P$ and $Pb_2(SO_4)O$ with a molar ratio of 1:1 were fully mixed and ground, and then the mixture powder was pressed into a cylinder with a diameter of 1.5 cm by a tablet press under a pressure of 100 MPa. Then the cylinder material was sealed in a quartz ampoule under a vacuum of $10^{-2}$ Pa, heated to 925 °C within 2 h and kept for 12 h. After the heat preservation, it was naturally cooled with the furnace.

The resistance-temperature (R-T) curves of as-synthesized LK-99 sample were measured by the standard four-probe method in a physical property measurement system equipped with BRT module (PPMS, DynaCool, Quantum Design). We conducted 25 times at each temperature point for average with a constant current mode. The cooling rate was set as 2 K min$^{-1}$ with the interval of 1 K. The magnetization-temperature (M-T) curves were measured with VSM module. The temperature sweeping rate was set as 2 K min$^{-1}$ and a 0.1 T magnetic field was applied.

**Results and Discussions**

**Figure 1** shows the X-ray diffraction (XRD) pattern of the synthesized LK-99 sample, which is very similar to that reported by Sukbae Lee et al.,[5, 6] proving that our synthesized material is LK-99. Additionally, there are other characteristic peaks that match well with the standard PDF cards of $Cu_2S$ and $Cu_2O$, indicating that the synthesized samples have impurities. **Figure 2a** is a photograph of the synthesized LK-99 sample, whose surface is rough and dark gray-black (**Figure 2b**). After polishing, the surface becomes obviously brighter (**Figure 2c, d**), we broke it and selected a small piece (sample 1) for the resistance-temperature (R-T) test, as shown in **Figure 2c**, the thickness of the sample is about 0.5 mm. Silver was used as the contact electrode, gold wire was used as the lead, and a 1 mm thick PCB was employed as the substrate to eliminate any stray current interference. **Figure 2e** shows the corresponding R-T curve. As can be seen, the LK-99 sample measured exhibits a typical semiconductor behavior on the whole, but there is a sudden resistance jump near 387 K (inset of **Figure 2e**). To verify that this jump phenomenon is not an accidental electrode contact fault, we repeatedly measured the R-T curve. As shown in **Figure 2f**, a steep resistance jump was still observed near 387 K, which confirmed the authenticity of the observed resistance jump results.

We speculate that there may be two explanations for this resistance jump phenomenon: one is that the material contains a superconducting phase component, which forms a connection with the semiconductor phase component. The second is that the material or impurity undergoes a previously unclear phase transition at this temperature. To further verify the above two conjectures, we conducted resistance-temperature tests under different magnetic fields applied. As shown in **Figure 3a**, the resistance jump temperature unchanged as the applied magnetic field increases, which is different from typical superconductors.[7-9] However, whether such a phase transition temperature that varies with the magnetic field also exists in semiconductors is still uncertain. To verify that this phenomenon is not an accidental phenomenon in a single sample, we also tested another LK-99 sample (sample 2), and the result still shows that the resistance jumps around 380 K (heating) and 364 K (cooling), and the jump temperature does not change with the magnetic field or applied current, as shown in **Figure 3b, c**.

Finally, the thermomagnetic curves (M-T, zero-field cooling (ZFC) and field cooling (FC) modes) of sample 1 in **Figure 2c** was also conducted, as shown in **Figure 3d**, where a distinct diamagnetic transition near 341 K and 354 K can be clearly observed. The temperature at the transition point is smaller than the resistance jump temperature, which different from the characteristics of room temperature superconductors. Considering the materials contain both $Pb_{10-x}Cu_x(PO_4)_6O$ and impurity identified at least including $Cu_2S$ and $Cu_2O$, it's difficult to conclude if such abnormal resistance behavior comes from LK-99 or impurity.

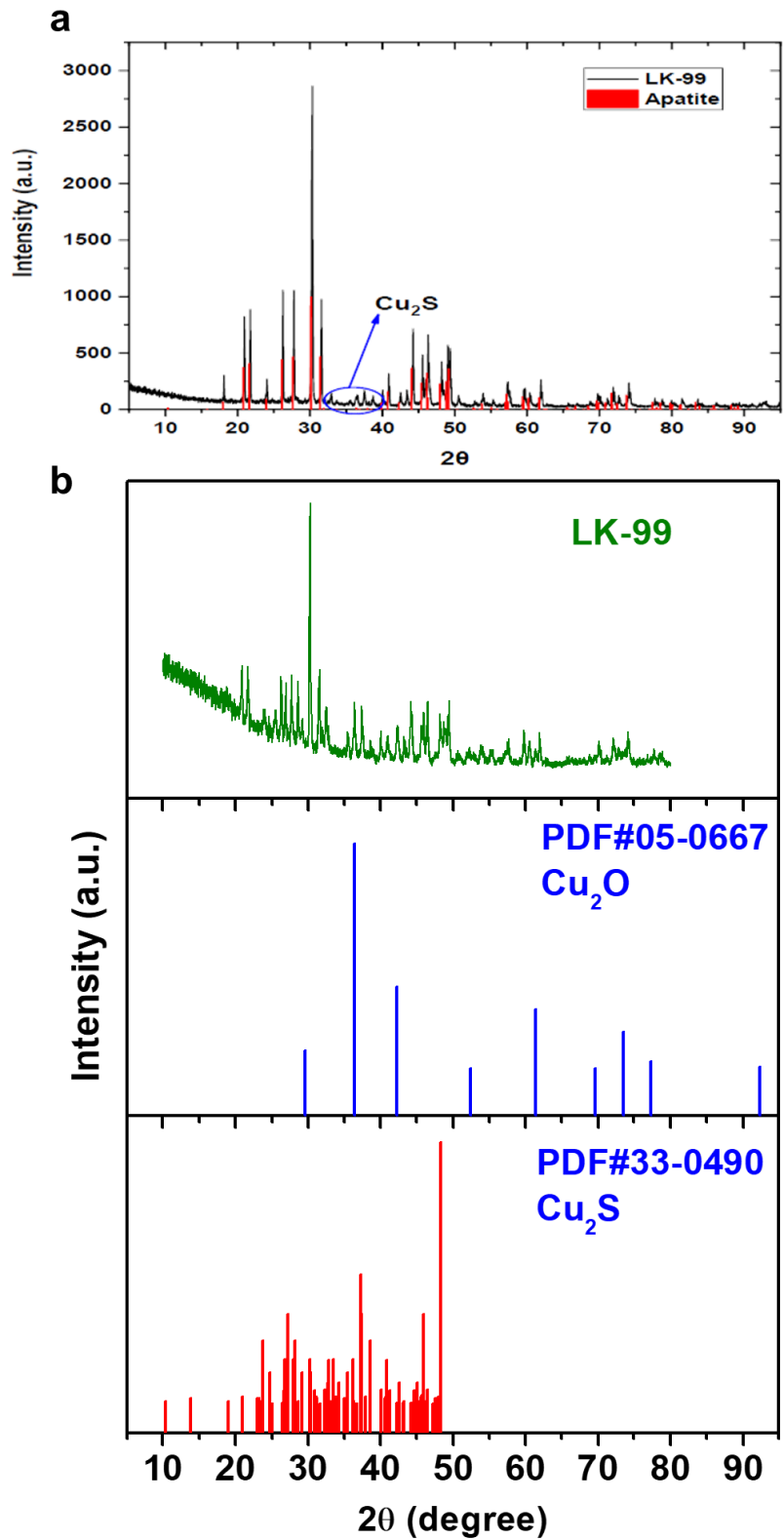

Figure 1. Impurity analysis of as-synthesized LK-99 sample. (a, b) XRD patterns of (a) ref. [5] and (b) LK-99 sample. The standard diffraction peak patterns of $Cu_2S$ and $Cu_2O$ from PDF database are shown for reference.

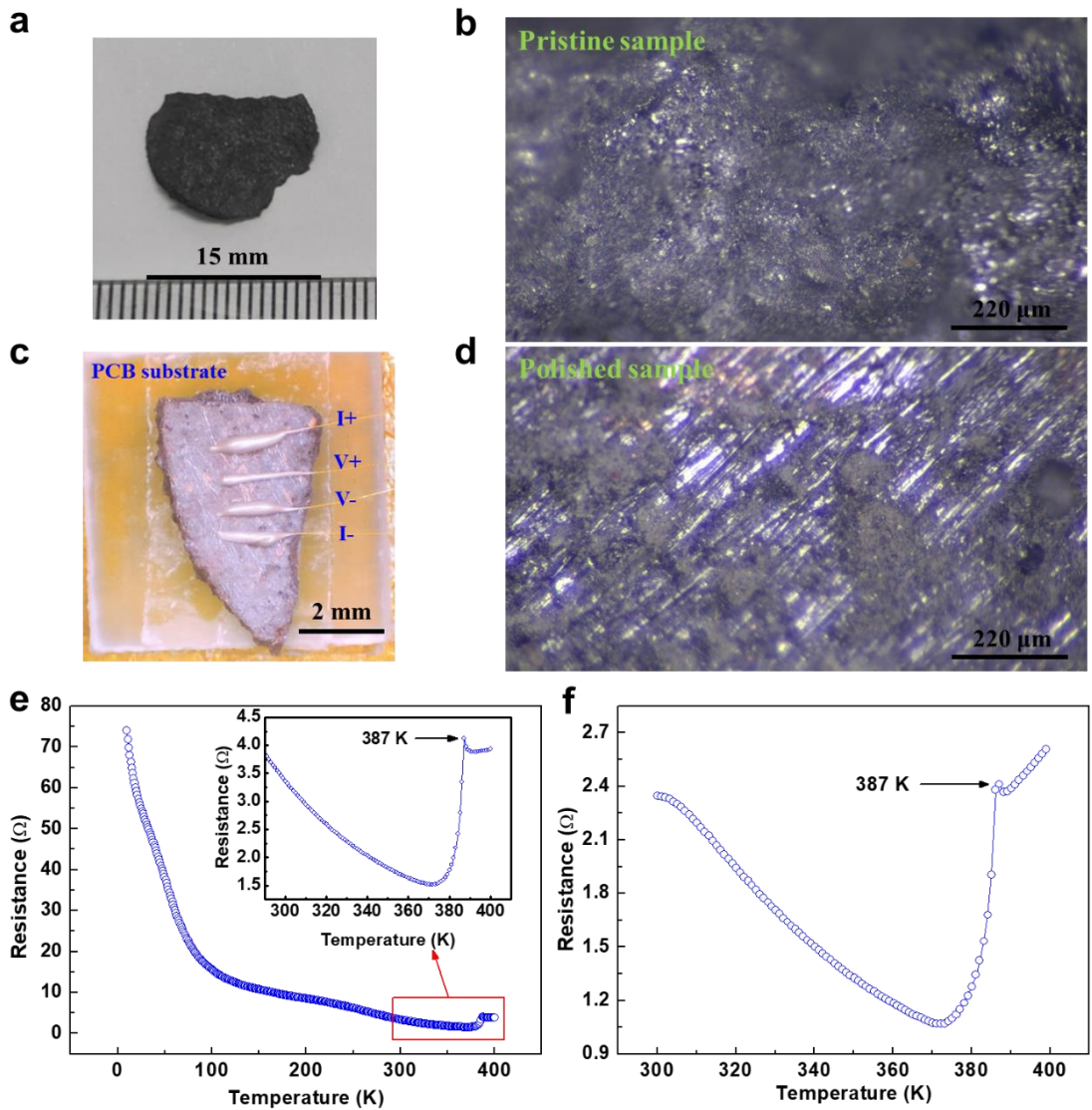

Figure 2. Synthesis and electrical transport property of LK-99 sample. (a, b) Photograph (a) and corresponding optical microscopic image (b) of as-synthesized LK-99 sample. (c, d) Photograph (c) and corresponding optical microscopic image (d) of as-synthesized LK-99 sample after polishing (sample 1). (e) Temperature-dependent resistance of the LK-99 sample in (c) in a temperature regime of 10-400 K. Magnetic field: 0 T. Inset shows an enlarged view of the temperature range of the red mark box, where a resistance jump could be clearly observed at ~387 K. (b) Repeated test of temperature-dependent resistance of the LK-99 sample in (c) in a temperature regime of 300-400 K. Magnetic field: 0 T.

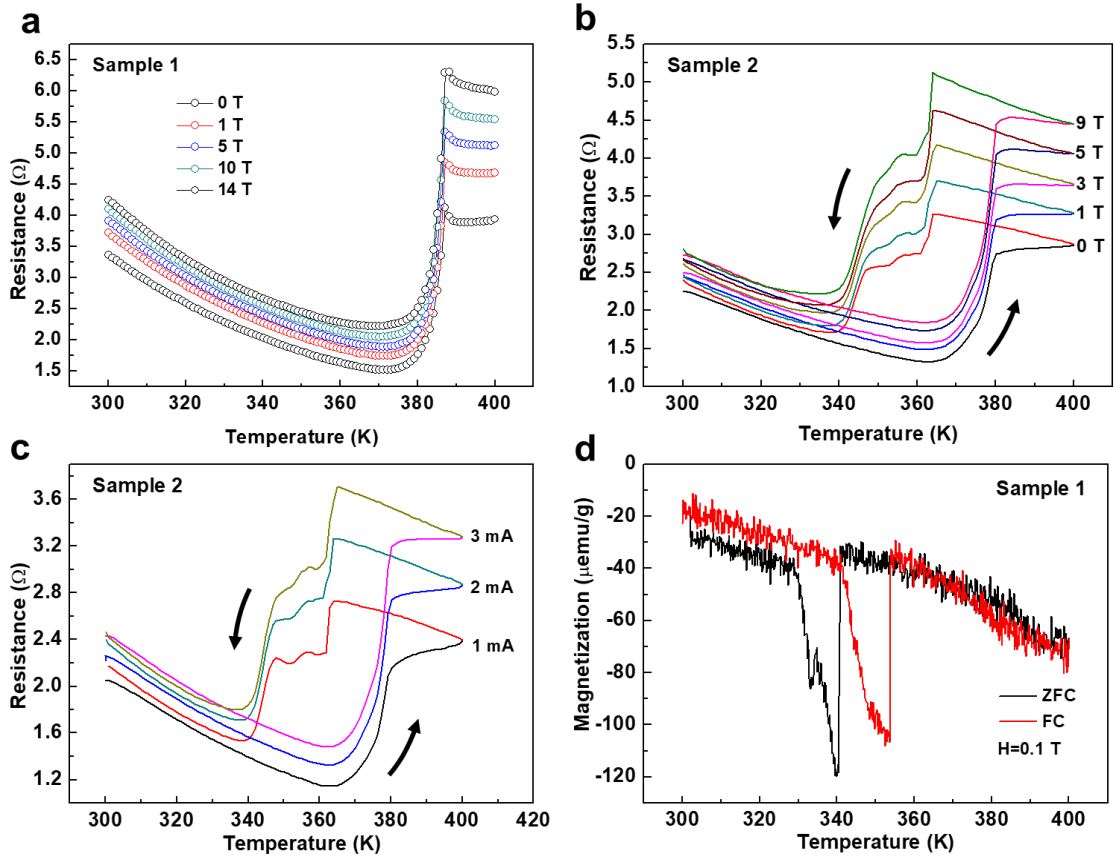

Figure. 3 Magnetoelectrical transport properties of LK-99 samples. (a, b) R-T curves of LK-99 sample 1 (a) and sample 2 (b) with different magnetic field applied. (c) R-T curves of LK-99 sample 2 with different applied current. (d) ZFC and FC curves of sample 1.

**Conclusion**

In short, the LK-99 sample we obtained is a mixed phase, in which the semiconductor phase is dominant. The overall temperature resistance characteristics show a typical semiconductor behavior, but an abnormal resistance jump is found near 387 K, which may come from $Pb_{10-x}Cu_x(PO_4)_6O$ or impurity such as $Cu_2S$ or $Cu_2O$. The resistance jump temperature keeps unchanged with the increase of the applied magnetic field. The diamagnetic jump temperature is slightly lower than the resistance jump temperature. To identify the intrinsic property of $Pb_{10-x}Cu_x(PO_4)_6O$, high quality crystals with high purity are highly needed. We should also note the possible contribution from impurity especially in resistance tests.

**Acknowledgments**

This work was supported by the National Key Research and Development Program of China (grant no. 2022YFE0134600), the National Natural Science Foundation of China (grant nos. 52272152, 61674063 and 62074061), the Natural Science Foundation of Hubei Province, China (grant no. 2022CFA031), the Foundation of Shenzhen Science and Technology Innovation Committee (grant nos. JCYJ20180504170444967 and JCYJ20210324142010030), and the fellowship of China Postdoctoral Science Foundation (grant no. 2022M711234).

**Competing interests**

## Data availability

The data that support this study are available from the corresponding authors upon reasonable request.